\documentstyle[aps,twocolumn,psfig]{revtex}

\title{Nonlinearities and Effects of Transverse Beam Size  
       in Beam Position Monitors (revised)}
\author{Sergey~S.~Kurennoy}
\address{Los Alamos National Laboratory,
         Los Alamos, NM 87545, USA}

\begin{document}
\maketitle

\begin{abstract}
The fields produced by a long beam with a given transverse charge 
distribution in a homogeneous vacuum chamber are studied. Signals 
induced by a displaced finite-size beam on electrodes of a beam 
position monitor (BPM) are calculated and compared to those produced
by a pencil beam. The non-linearities and corrections to BPM signals 
due to a finite transverse beam size are calculated for an arbitrary 
chamber cross section. Simple analytical expressions are given for a 
few particular transverse distributions of the beam current in a 
circular or rectangular chamber. Of particular interest is a general 
proof that in an arbitrary homogeneous chamber the beam-size 
corrections vanish for any axisymmetric beam current distribution.
\end{abstract}

\section{Introduction} 

In many accelerators, especially in ion linacs and storage rings, 
beams occupy a significant fraction of the vacuum chamber cross 
section. On the other hand, an analysis of beam-induced signals 
in beam position monitors (BPMs) is usually restricted to the case 
of an infinitely small beam cross section (pencil beam). In this 
paper we consider the problem for a vacuum chamber with an arbitrary 
but constant cross section, and calculate, for a given transverse 
charge distribution of an off-axis relativistic beam, the fields 
produced by the beam on the chamber wall. Comparing those with the 
fields of a pencil beam gives us corrections (e.g., to BPM signals) 
due to a finite transverse size of the beam.

Let a vacuum chamber have an arbitrary single-connected 
cross section $S$ that does not change as a beam moves along the 
chamber axis $z$, and perfectly conducting walls. We consider the 
case of $(\omega b/\beta \gamma c)^2 \ll 1$, where $\omega$ is the 
frequency of interest, $\beta c$ is the beam velocity, 
$\gamma=1/\sqrt{1-\beta^2}$, and $b$ is a typical transverse 
dimension of the vacuum chamber. It includes both 
the ultra relativistic limit, $\gamma \gg 1$, and 
the long-wavelength limit when, for a fixed $\gamma$, 
the wavelength of interest $\lambda \gg 2\pi b/\gamma$.
Under these assumptions the problem of calculating the 
beam fields at the chamber walls is reduced to a 2-D electrostatic 
problem of finding the field of the transverse distribution 
$\lambda(\vec{r})$ of the beam charge, which occupies region $S_b$ 
of the beam cross section, on the boundary $\partial S$ of region 
$S$, see e.g.\ \cite{Cuperus}. The layout of the problem is 
illustrated in Fig.\ 1.

Let the beam charge distribution $\lambda(\vec{r})$ satisfy the 
normalization condition $\int_{S_b} d\vec{r} \lambda(\vec{r})=1$, 
which means the unit charge per unit length of the bunch. If we know 
the field $e(\vec{r},\vec{b})$ produced at a point $\vec{b}$ on the 
wall by a pencil beam located at a point $\vec{r}$ of region $S_b$, 
the field of the distribution is given by
\begin{equation}
E(\vec{a},\vec{b}) = \int_{S_b} d\vec{r} \lambda(\vec{r}) 
 e(\vec{r},\vec{b}) \ ,                                \label{fldd}
\end{equation}
where the vector $\vec{a}$ is defined as the center of the charge 
distribution: $\vec{a} = \int d\vec{r} \vec{r} \lambda(\vec{r})$. 
Obviously, the case of a pencil beam corresponds to the 
distribution  $\lambda(\vec{r})=\delta(\vec{r}-\vec{a})$, where 
$\delta(\vec{r})$ is the 2-D $\delta$-function. 
Let us start from a particular case of a circular cylindrical 
vacuum chamber.

\section{Circular Chamber}

In a circular cylindrical pipe of radius $b$, a pencil beam with a 
transverse offset $\vec{r}$ from the axis produces the following 
electric field on the wall
\begin{eqnarray}
e(\vec r,\vec b) & = & \frac{1}{2\pi b} \frac{b^2 - r^2}
 {b^2 - 2br\cos (\theta - \varphi ) + r^2 }              \label{fld0}
 \\ & = &
 \frac{1}{{2\pi b}}\left\{ {1 + 2\sum\limits_{k = 1}^\infty 
 {\left( {\frac{r}{b}} \right)^k \cos } \left[ {k(\theta 
  - \varphi )} \right]} \right\} , \nonumber
\end{eqnarray}
where $\varphi,\theta$ are the azimuthal angles of vectors 
$\vec r,\vec b$, correspondingly. One should note that this field 
is normalized as follows:
$$
 \oint_{\partial S}\! dl \ e(\vec r,\vec b) = 1 \ , 
$$
where integration goes along the boundary ${\partial S}$ of the 
transverse cross section of the vacuum chamber. 

Integrating the multipole expansion in the RHS of Eq.\ (\ref{fld0}) 
with a double-Gaussian distribution of the beam charge
\begin{equation}
\lambda (x,y) = \frac{1}{2\pi \sigma _x \sigma _y }
 \exp \left[  - \frac{{\left( {x - a_x } \right)^2 }}
 {{2\sigma _x^2 }} - \frac{{\left( {y - a_y } \right)^2 }}
 {{2\sigma _y^2 }} \right] \ ,                           \label{dblg}
\end{equation}
--- assuming, of course, that the rms beam sizes are small,
$\sigma_x,\sigma_y \ll b$, 
--- one obtains non-linearities in the form of powers of $a_x,a_y$, 
as well as the beam size corrections, which come as powers of 
$\sigma_x,\sigma_y$. 
To our knowledge, this was done first for the double-Gaussian beam 
in a circular pipe by R.H. Miller {\em et al} in the 1983 paper 
\cite{Miller}, where the expansion was calculated up to the 3rd 
order terms. More recently, their results have been used at LANL 
in measuring second-order beam moments with BPMs and calculating 
the beam emittance from the measurements \cite{R&C}. 
In a recent series of papers \cite{CERN} by CERN authors, 
the results \cite{Miller} have been recalculated (and corrected in 
the 3rd order), and used to derive the beam size from measurements 
with movable BPMs. 

In fact, integrating (\ref{fld0}) with the distribution (\ref{dblg}) 
can be readily carried out up to an arbitrary order. Using in 
Eq.\ (\ref{fld0}) the binomial expansion for
$$
r^k \cos \left[ {k(\theta  - \varphi )} \right] = 
 {\mathop{\rm Re}\nolimits} \left[ {e^{ik\theta } 
 \left( {x - iy} \right)^k } \right]
$$
makes the $x$- and $y$-integrations very simple, and the $k$-th order 
term ($k$-pole) of the resulting expansion is 
\begin{eqnarray}
E^{(k)} (\theta ) & = & 
 \frac{{k!}}{\pi b}\sum\limits_{m = 0}^k {\cos } 
  \left( {\frac{m\pi }{2} - k\theta } \right)
  \frac{x_0^{k - m} y_0^m }{b^k } \nonumber \\
  & \times & \sum\limits_{s = 0}^{[(k - m)/2]} 
  {\frac{{(\sigma _x^2 /2x_0^2 )^s }}{{s!(k - m - 2s)!}}} 
  \sum\limits_{p = 0}^{[m/2]} 
  {\frac{(\sigma _y^2 /2y_0^2 )^p }{p!(m - p)!}} ,     \label{kcirc}
\end{eqnarray}
where $x_0,y_0$ stand for the beam center coordinates $a_x,a_y$. 
Explicitly, up to the 5th order terms,
\begin{eqnarray}
\lefteqn{E(\vec r_0 ,\vec b) = \frac{1}{2\pi b} + \frac{1}{\pi b^2}
 \left\{ \cos \theta\;x_0 + \sin \theta\;y_0 \right\} } \nonumber
\\ \lefteqn{ + \frac{1}{\pi b^3}\left\{ {\cos 2\theta \;
  \left( {\sigma _x^2  - \sigma _y^2  + x_0^2  - y_0^2 } \right) 
  + \sin 2\theta \;2x_0 y_0 } \right\} } \nonumber
\\ & & + \frac{1}{\pi b^4}\left\{ \begin{array}{l}
 \! \cos 3\theta \;x_0 \left[ {3\left( {\sigma _x^2 - \sigma _y^2 } 
   \right) + x_0^2  - 3y_0^2 } \right] +   \\ 
 \! \sin 3\theta \;y_0 \left[ {3\left( {\sigma _x^2  - \sigma _y^2 } 
   \right) + 3x_0^2  - y_0^2 } \right]  \\ 
 \end{array}  \! \! \right\}                           \label{k5circ}
\\ \lefteqn{ + \frac{1}{\pi b^5}\left\{ \begin{array}{l}
 \! \cos 4\theta \;\left[ \begin{array}{l}
 3\left( \sigma _x^2 - \sigma _y^2 + x_0^2 - y_0^2 \right)^2  \\ 
  - 2x_0^4  - 2y_0^4   \\ 
 \end{array} \right] +  \\ 
 \! \sin 4\theta \;4x_0 y_0 \left[ {3\left( \sigma _x^2 - 
   \sigma _y^2 \right) + x_0^2  - y_0^2 } \right]  \\ 
 \end{array} \! \! \right\} } \nonumber
\\ \lefteqn{ + \frac{1}{\pi b^6}\left\{ \begin{array}{l}
  \! \cos 5\theta \;x_0 \left[ \begin{array}{l}
 15\left( {\sigma _x^2  - \sigma _y^2 } \right)^2  +   \\ 
 10\left( {\sigma _x^2  - \sigma _y^2 } \right)\left(x_0^2 - 3y_0^2
    \right) \\  + x_0^4  - 10x_0^2 y_0^2  + 5y_0^4   \\ 
 \end{array} \right] \! + \!  \\ 
 \! \sin 5\theta \;y_0 \left[ \begin{array}{l}
 15\left( {\sigma _x^2  - \sigma _y^2 } \right)^2  +  \\ 
 10\left( {\sigma _x^2  - \sigma _y^2 } \right)\left(3x_0^2 - y_0^2
  \right)  \\ 
   + 5x_0^4  - 10x_0^2 y_0^2  + y_0^4  \\ 
 \end{array} \right]  \\ 
 \end{array} \! \! \right\} } \nonumber
\\ \lefteqn{ +  \ldots }  \nonumber
\end{eqnarray}
The multipole expansion (\ref{k5circ}) that includes terms up to 
the decapole ones, leads us to one interesting observation: all 
beam-size corrections come in the form of the difference 
$\sigma _x^2-\sigma _y^2$, and vanish for a round beam where 
$\sigma _x^2=\sigma _y^2$. This would be obvious for an on-axis beam 
in a round pipe from the Gauss law, but for a deflected beam the 
result seems rather remarkable. 

It is not easy to see directly from Eq.\ (\ref{kcirc}) whether the 
beam-size corrections for a round beam in a round pipe vanish in all 
orders. However, one can check explicitly that it is the case. Let us 
consider an arbitrary azimuthally symmetric distribution of the beam 
charge $\tilde{\lambda}(\vec{r})=\tilde{\lambda}(r)$, where the tilde 
in $\tilde{\lambda}$ means that the argument of the 
distribution function $\lambda$ is shifted so that the vector 
$\vec{r}$ now originates from the beam center: 
$\lambda(\vec{a}+\vec{r})=\tilde{\lambda}(\vec{r})$.

In this case, the integration in Eq.\ (\ref{fldd}) for the circular 
pipe can be done explicitly. Namely, using the expansion in 
(\ref{fld0}) and integrating in polar coordinates $(r,\varphi)$, 
for the case when $\tilde{\lambda}(r,\varphi)=\tilde{\lambda}(r)$ 
one can write
\begin{eqnarray}
\lefteqn{ \qquad E(\theta ) =  \frac{1}{2\pi b} 
  \int\limits_0^\infty rdr \tilde \lambda (r) \times } \nonumber
 \\ & & \int\limits_0^{2\pi} 
 \frac{d\varphi \left( b^2 - a^2 - r^2 - 2br\cos \varphi \right)}
 {b^2 + a^2 + r^2 - 2ab\cos \theta + 2ar\cos \varphi 
      - 2br\cos (\varphi  - \theta )}  \nonumber    
  \\ \lefteqn{ \qquad =  \frac{1}{2\pi b}\int\limits_0^\infty 
  2\pi rdr \tilde \lambda (r)\frac{b^2 - a^2}
  {b^2 + a^2 - 2ab\cos \theta } }                        \label{icirc}
 \\ \lefteqn{ \qquad = \frac{1}{2\pi b}\frac{b^2 - a^2}
    {b^2 + a^2 - 2ab\cos \theta } \ . } \nonumber   
\end{eqnarray}
The last expression follows from the preceding one due to the charge 
normalization, and it is exactly the field of a pencil beam displaced 
from the chamber axis by $\vec a=(a,0)$, compare Eq.\ (\ref{fld0}). 
The only real effort here was to perform the angular integration, 
which turns out to be independent of $r$. It was done analytically 
by introducing a new complex variable $z$: $\cos \varphi = 
(z+z^{-1})/2$, and then integrating along a unit circle in the 
complex $z$-plane with residues\footnote{Trying to perform 
this integration with {\it Mathematica}, I found a bug in its 
analytical integration package for this particular kind of 
integrals. Wolfram Research acknowledged the bug, and they work 
to fix it.}.

Now we apply the above results for calculating signals in a BPM.
First of all, we will assume that signals induced in BPM electrodes 
(striplines or buttons) are proportional to the wall image current 
integrated within the transverse extent of the electrode on the 
chamber wall. Such an assumption is usually made in analytical 
treatments of BPM signals, see e.g.\ \cite{Cuperus,Miller,CERN,RS89}, 
and is justified when the BPM electrodes are flush with the chamber 
walls, grounded, and separated from the wall by narrow gaps. 
Certainly, there are some field distortions due to the presence of 
the gaps, but they are rather small for narrow gaps. Moreover, even 
for a more complicated BPM geometry with realistic striplines 
protruding inside a circular pipe, it was demonstrated by 
measurements (see in \cite{RS89}) and by numerical 3-D modeling 
\cite{SK00} that the effects of field distortions near the stripline 
edges can be accounted for by integrating the wall current within an 
effective transverse extent of the striplines (slightly larger than 
their width) in a simple smooth-pipe model with the effective pipe 
radius taken to be an average of the stripline inner radius and 
that of the beam pipe.  

Consider now in a circular chamber of radius $b$ a stripline BPM 
with a pair of electrodes in the horizontal plane. Let us assume 
that the stripline electrodes are flush with the chamber walls,
grounded, and have subtended angle $\phi$ per stripline. Following 
the discussion above, we neglect the field distortions near the 
strip edges, and calculate the signals induced on the stripline 
electrodes by integrating the field (\ref{k5circ}) over the interval 
$-\phi/2 \le \theta \le \phi/2$ for the right electrode ({\it R}) 
and over $\pi-\phi/2 \le \theta \le \pi+\phi/2$ for the left one 
({\it L}). The ratio of the difference between the signals on 
the right and left electrodes in the horizontal plane to the sum 
of these signals is 
\begin{eqnarray}
\lefteqn{ \frac{R - L}{R + L} =
 2\frac{x_0}{b}\frac{\sin \phi /2}{\phi /2} 
  \times \left\{\; 1 \; - \right. } \nonumber
\\ && - \frac{2}{b^2}\frac{{\sin \phi }}{\phi }
 \left( \sigma_x^2 - \sigma_y^2 + x_0^2 - y_0^2 \right) \nonumber 
\\ && + \frac{1}{b^2}\frac{\sin 3\phi /2}{\sin \phi /2} \left( 
 \sigma_x^2  - \sigma_y^2  + x_0^2 /3 - y_0^2 \right) \label{ratcirc}
\\&& - \frac{2}{b^4}\frac{\sin \phi }{\phi }
   \frac{\sin 3\phi /2}{\sin \phi /2} \left( \sigma_x^2 
     - \sigma_y^2 + x_0^2 - y_0^2 \right) \times    \nonumber
\\&& \qquad \times \left( \sigma_x^2 
      - \sigma_y^2 + x_0^2 /3 - y_0^2 \right)      \nonumber   
\\&& - \frac{2}{b^4}\frac{\sin 2\phi}{2\phi }
 \left[ 3 {\left( {\sigma_x^2 - \sigma_y^2 + x_0^2 - y_0^2}\right)^2
   - 2x_0^4  - 2y_0^4 } \right]  \nonumber
\\ && + \frac{4}{b^4}\left( \frac{\sin \phi}
   {\phi} \right)^2 \left( \sigma _x^2 - \sigma _y^2
    + x_0^2 - y_0^2 \right)^2  \nonumber
\\ && + \frac{1}{b^4}\frac{{\sin 5\phi /2}}
 {{\sin \phi /2}}\left[ \begin{array}{l}
 3\left( {\sigma _x^2  - \sigma _y^2 } \right)^2  \\ 
  + 2\left( {\sigma _x^2  - \sigma _y^2 } \right)
  \left( {x_0^2  - 3y_0^2 } \right) \\ 
  + x_0^4 /5 - 2x_0^2 y_0^2  + y_0^4  \\
 \end{array} \right] \nonumber 
\\ && \left. + O\left( b^{- 6} \right) \right \} \ . \nonumber
\end{eqnarray} 
The factor outside the brackets in the RHS of Eq.\ (\ref{ratcirc}) 
is the linear part of the BPM response, so that all terms in the 
brackets except 1 are non-linearities and beam-size corrections.

Corrections (\ref{ratcirc}) are shown in Figs.\ 2-5 for a 
60${}^\circ$ stripline BPM. Figure 2 shows the non-linearities 
of the BPM response for a pencil beam. The changes of this signal 
ratio when the beams become flat (Figs.\ 3-4) are practically 
unnoticeable. In Fig.\ 5 the ratio $S/S_0$, where 
$S=(R - L)/(R + L)$ for a finite-size beam, and $S_0$ is the 
same difference-over-sum ratio but for a pencil beam, is plotted 
versus the beam center position. One can conclude from Fig.\ 5 
that in this BPM for a reasonable transverse beam size the 
beam-size corrections are on the level of one percent. One can 
check that the weak dependence of the signal ratio on the beam 
size and on the vertical beam offset $y_0$ in this case is mainly 
due to the fact that the BPM electrodes are wide. For narrow 
electrodes the beam-size and transverse coupling effects are 
stronger, on the order of a few percents, see in Sect.\ IV.   

\section{Vacuum Chamber of Arbitrary Cross Section}

Let us consider now a more general case of a homogeneous vacuum 
chamber with an arbitrary single-connected cross section $S$. 
The field $e(\vec{r},\vec{b})$ produced by a pencil beam at a point 
$\vec{b}$ on the wall can be written as (see e.g.\ \cite{KGS,SK96}) 
\begin{equation}
e(\vec r,\vec b) =  - \sum\nolimits_s {k_s^{ - 2} } 
e_s (\vec r)\nabla _\nu  e_s (\vec b) \ ,               \label{gexp}
\end{equation}
where $s=(n,m)$ is a generalized (2-D) index,   
$\nabla _\nu = \vec{\nabla} \cdot \vec{\hat{\nu}}$
is a normal derivative at the point $\vec{b}$ on the region 
boundary $\partial S$ ($\vec \nabla$ is the 2-D gradient operator,
$\vec{\hat{\nu}}$ means an outward normal vector to the boundary), 
and $k_s^2,e_s (\vec r)$ are eigenvalues and orthonormalized 
eigenfunctions of the following 2-D Dirichlet problem in the 
region $S$:
\begin{equation}
\left( {\nabla ^2  + k_s^2 } \right)e_s (\vec r) = 0;
 \quad e_s (\vec r \in \partial S) = 0 \ .        \label{bound}
\end{equation}

The expansion (\ref{gexp}) follows from the fact that 
\begin{equation}
\Phi (\vec r - \vec a) = \sum\nolimits_s {k_s^{ - 2} } 
 e_s (\vec r)e_s (\vec a) \ .                      \label{fgreen}
\end{equation}
is the Green function of the problem (\ref{bound}), 
which means that it satisfies the equation
\begin{equation}
\nabla ^2 \Phi (\vec r - \vec a) =  
- \delta (\vec r - \vec a)  \ .                      \label{equfgr}
\end{equation}
In other words, $\Phi (\vec r - \vec a)$ is (up to a factor 
$1/\varepsilon_0$) an electric potential created at point $\vec r$ 
of region $S$ by a unit point charge placed at point $\vec a$. 
One can easily check that substituting the sum (\ref{fgreen}) 
into Eq.\ (\ref{equfgr}) gives, with the account of (\ref{bound}), 
the correct result due to the following property of eigenfunctions
\begin{equation}
\sum\nolimits_s {e_s (\vec r)e_s (\vec a)}  
= \delta (\vec r - \vec a) \ .                        \label{eigpro}
\end{equation}
The eigenfunctions for simple regions like a circle or a rectangle 
can be found in electrodynamics textbooks (or see Appendix in 
Ref.\ \cite{KGS}). For the circular case, summing the corresponding 
Bessel functions in (\ref{gexp}) leads directly to the last 
expression in Eq.\ (\ref{fld0}).

For a thick beam with a given transverse charge distribution, one 
can write from Eqs.\ (\ref{fldd}) and (\ref{gexp}):
\begin{equation}
E(\vec a,\vec b) =  - \sum\nolimits_s {k_s^{ - 2} } 
 \nabla _\nu  e_s (\vec b)\int_{S_b } {\tilde \lambda 
 (\vec r)e_s (\vec a + \vec r)d\vec r} \ ,              \label{gfld}
\end{equation}
where again the tilde in $\tilde{\lambda}$ means an argument shift 
in the distribution function $\lambda$: 
$\lambda(\vec{a}+\vec{r})=\tilde{\lambda}(\vec{r})$, 
so that the integration vector $\vec{r}$ originates from the beam 
center $\vec{a}$. Performing the Taylor expansion of the 
eigenfunction $e_s (\vec a + \vec r)$ around point $\vec{a}$
\begin{eqnarray*}
e_s (\vec a + \vec r) 
 & = & \sum\limits_{m = 0}^\infty 
  {\left( {\vec r\vec \nabla } \right)^m e_s (\vec a)} /m! \\
 & = & e_s (\vec a) + \vec r\vec \nabla e_s (\vec a) + 
 \frac{1}{2}\left( {\vec r\vec \nabla } \right)^2 
 e_s (\vec a) +  \ldots  
\end{eqnarray*}
and integrating in (\ref{gfld}) leads to the following multipole 
series:
\begin{eqnarray}
\lefteqn{ E(\vec a,\vec b) = - \sum\nolimits_s k_s^{- 2} 
 \nabla_\nu e_s (\vec b) \sum\limits_{m = 0}^\infty
 \sum\limits_{i_1 = 1}^2 \sum\limits_{i_2 = 1}^2 
  \ldots }                                              \label{mexp}
\\ && \times \sum\limits_{i_m = 1}^2 \partial_{i_1 } 
  \partial_{i_2} \ldots \partial_{i_m } e_s (\vec a) /m!  
  \int_{S_b } d\vec r\tilde \lambda (\vec r) 
   r_{i_1} r_{i_2} \ldots r_{i_m }  \nonumber \ ,  
\end{eqnarray}
where $\partial _i  \equiv \partial /\partial r_i ,\;i = 1,2$, 
and all effects of the finite beam size here enter through the 
components of the multipoles of the beam charge distribution.

If we restrict ourselves by considering only symmetric 
(with respect to two axis) charge distributions, 
i.e.\ assume $\tilde{\lambda}(-\vec{r})=\tilde{\lambda}(\vec{r})$, 
all integrals for odd $m$s in (\ref{mexp}) vanish, 
and the general expansion (\ref{mexp}) can be significantly 
simplified:
\begin{eqnarray}
\lefteqn{ E(\vec a,\vec b) = e(\vec a,\vec b) + 
 \frac{1}{2}\partial _x^2 e(\vec a,\vec b)\int_{S_b} d\vec r
  \tilde \lambda (\vec r)\left( x^2 - y^2 \right) } \nonumber  
\\ && \ + \frac{1}{24}\partial _x^4 e(\vec a,\vec b)
  \int_{S_b}d\vec r\tilde \lambda (\vec r)
  \left( x^4 - 6x^2 y^2 + y^4 \right) + \ldots          \label{mexp2}
\\ && \qquad = e(\vec a,\vec b) 
 + \sum\limits_{n = 1}^\infty \partial_x^{2n} 
 e(\vec a,\vec b) \; M_{2n} /\left( {2n} \right)! \nonumber
\end{eqnarray}
In obtaining the last expression the following property of 
the sum (\ref{mexp}) was used: flipping the derivatives like this
$\partial_y^2 e_s(\vec a) = - \partial_x^2 e_s(\vec a)$ inside the 
sum does not change the result. This is due to 
$\partial_y^2 e_s(\vec a) = - \left( \partial_x^2+k_s^2 \right )
 e_s(\vec a)$ from Eq.\ (\ref{bound}), and because 
any extra factor $k_s^2$ in (\ref{mexp}) leads to a zero sum 
since it just gives a derivative of the $\delta$-function, cf.\
(\ref{eigpro}), with a non-zero argument because of 
$\vec a \ne \vec b$ (the beam does not touch the wall). 

Equation (\ref{mexp2}) is more transparent than (\ref{mexp}). 
Let us take a look at the moments in (\ref{mexp2}) in their 
closed form:
\begin{eqnarray}
 M_{2n} =  \int_{S_b} d\vec r\tilde \lambda (\vec r)    \label{m2n}
  \left[ x^{2n} - C_2^{2n - 2} x^{2n - 2} y^2 + \right. 
 \\ \left. + C_4^{2n - 4} x^{2n - 4} y^4 - \ldots 
    + \left( {-1} \right)^n y^{2n} \right] \nonumber
\end{eqnarray}
where $C_k^n  = n!/\left[ {k!\left( {n - k} \right)!} \right]$
are binomial coefficients. 
It is useful to notice that the sum inside the square brackets in 
(\ref{m2n}) is simply ${\mathop{\rm Re}\nolimits} 
\left[ \left( {x + iy} \right)^{2n} \right] $, 
and in the polar coordinates of the beam Eq.\ (\ref{m2n}) 
can be rewritten simply as
\begin{equation}
M_{2n} = \int_{S_b }                                   \label{m2npol}
 d\vec r\tilde \lambda (\vec r) \;r^{2n} \cos 2n\varphi \ .
\end{equation}
Now it is quite obvious that if one assumes an arbitrary azimuthally 
symmetric distribution of the beam charge
$\tilde{\lambda}(\vec{r})=\tilde{\lambda}(r)$, 
i.e.\ $\tilde{\lambda}(r,\varphi)=\tilde{\lambda}(r)$, 
all beam moments (\ref{m2npol}) become equal to zero after the 
angular integration, and the corresponding beam-size corrections 
in (\ref{mexp2}) vanish. Therefore, we proved a rather general 
statement: {\it the fields produced by an ultra relativistic beam 
with an azimuthally-symmetric charge distribution on the walls of 
a homogeneous vacuum chamber of an arbitrary cross section are 
exactly the same as those due to a pencil beam of the same current 
following the same path}. 
A particular case of this statement, for a circular chamber cross 
section, was proved by explicit calculations earlier, in Sect.\ II.

The physical explanation of this effect is simple. The electric 
field outside the beam $\vec{E}$ is a superposition of the field
due to the charge distribution itself, $\vec{E}^{dis}_{vac}$, and 
the field due to induced charges on the chamber walls, 
$\vec{E}_{ind}$. From the Gauss law, for an azimuthally-symmetric 
beam charge distribution, the field $\vec{E}^{dis}_{vac}$ 
outside the beam (in vacuum, without the chamber) is exactly the 
same as that of a pencil beam, $\vec{E}^0_{vac}$, if the last one 
has the same charge and travels along the axis of the thick beam. 
Therefore, the induced charge distribution on the wall will be 
identical for the thick and pencil beams, and as a result the 
same will be true for the total electric field outside the 
beam\footnote{This remark is due to a discussion with 
M. Blaskiewicz.}. 

The expansion (\ref{mexp2}) for symmetric distributions 
of the beam charge gives the beam-size corrections for an arbitrary 
chamber, as long as the beam charge distribution is known. 
As two particular symmetric charge distributions of practical 
interest, we consider a double Gaussian one, cf.\ Eq.\ (\ref{dblg}),
\begin{equation}
\tilde \lambda (x,y) = \exp \left( { - x^2 /2\sigma _x^2 
  - y^2 /2\sigma _y^2 } \right) /                      \label{dblgs}
  \left( {2\pi \sigma _x \sigma _y } \right) \ ,
\end{equation}
and a uniform beam with a rectangular cross section  
$2 \sigma _x \times 2 \sigma _y$
\begin{equation}
\tilde \lambda (x,y) = \frac{\theta \left( x + \sigma_x \right)
 \theta \left(\sigma_x - x \right) \theta \left(y+\sigma_y\right)
 \theta\left(\sigma_y-y\right)}{4\sigma_x\sigma_y}\ ,   \label{unifr}
\end{equation}
where $\theta(x)$ is the step function. The two distributions 
$\tilde \lambda $ above are written in the beam coordinates,
with $x=y=0$ corresponding to the beam center, as was discussed
after Eq.\ (\ref{k5circ}).

For the double Gaussian beam (\ref{dblgs}), 
$M_2 = \sigma _x^2 - \sigma _y^2$,  
$M_4 = 3\left( {\sigma _x^2 - \sigma _y^2 } \right)^2$, etc., 
so that from Eq.\ (\ref{mexp2}) follows
\begin{eqnarray}
 E(\vec a,\vec b) & = & e(\vec a,\vec b) + \frac{1}{2}
 \left( {\sigma _x^2  - \sigma _y^2 } \right)\partial _x^2 
 e(\vec a,\vec b)  \nonumber   
\\ & + & \frac{1}{8} \left( {\sigma_x^2 - \sigma_y^2} \right)^2 
 \partial_x^4 e(\vec a,\vec b)                          \label{mexpg}
\\ & + & \frac{1}{48} \left( {\sigma _x^2 - \sigma _y^2 } \right)^3 
 \partial _x^6 e(\vec a,\vec b) + O(\sigma ^8 ) \ . \nonumber   
\end{eqnarray}

Similarly, for the uniform beam with the rectangular cross section 
(\ref{unifr}), the corrections are 
\begin{eqnarray}
E(\vec a,\vec b) & = & e(\vec a,\vec b) + \frac{1}{6}
 \left( {\sigma _x^2  - \sigma _y^2 } \right)\partial _x^2 
 e(\vec a,\vec b) \nonumber
\\& + & \frac{1}{40} \left( {\sigma_x^4 - 
   \frac{10}{3} \sigma _x^2 \sigma_y^2  + \sigma_y^4 } \right)
 \partial _x^4 e(\vec a,\vec b)                         \label{mexpu}
\\& + & \frac{1}{5040} \left( \sigma_x^6 - 7\sigma_x^4 
 \sigma_y^2 + 7\sigma_x^2 \sigma_y^4 - \sigma_y^6 \right)
 \partial_x^6 e(\vec a,\vec b) \nonumber 
\\ & + & O(\sigma ^8 ) \ .  \nonumber
\end{eqnarray}

One can see that for a round beam, $\sigma _x = \sigma _y$, 
all corrections in (\ref{mexpg}) disappear as expected, and for a 
square beam cross section in (\ref{mexpg}), the lowest correction is 
proportional to $\sigma^4$, while the next-order one to $\sigma^8$.

One should note at this point that the general field expansion
(\ref{mexp2}) and Eqs.\ (\ref{mexpg}-\ref{mexpu}) derived above 
are essentially the expansions in a small parameter $\sigma^2/b^2$, 
where $\sigma$ is a typical transverse beam size, and $b$ stands for 
a characteristic transverse dimension of the chamber cross section. 
The powers of $1/b$ are produced by the derivatives of the pencil 
beam field $e(\vec a,\vec b)$ in Eqs.\ (\ref{mexp2}) and 
(\ref{mexpg}-\ref{mexpu}). Therefore, these results are valid for 
any beam offset $a$, large or small, no matter what is the relation 
between $\sigma$ and $a$.

Equations (\ref{mexp2}) and (\ref{mexpg}-\ref{mexpu}) 
give us a rather good idea about how the beam-size corrections enter 
into the field expressions. The non-linearities, however, are hidden 
in the pencil-beam field $e(\vec a,\vec b)$ and in its 
derivatives. We can single out the non-linearities in a way similar 
to the one used to obtain the beam-size corrections, by expanding 
the field $e(\vec a,\vec b)$ in powers of $a$ around the chamber 
axis:
\begin{eqnarray*}
e(\vec a,\vec b) & = & \sum\limits_{m = 0}^\infty  
{\left( {\vec a\vec \nabla } \right)^m e(0,\vec b)} /m! \\
 & = & e_0  + \vec a\vec \nabla e_0  + \frac{1}{2}
\left( \vec a\vec \nabla \right)^2 e_0  +  \ldots \ ,
\end{eqnarray*}
where the notation $e_0=e(0,\vec b)$ was introduced for brevity, 
and similarly for the derivatives. In the most general case, 
unfortunately, it does not lead to convenient equations. However, 
for vacuum chambers with some symmetry the results can be 
simplified significantly. Here we limit our consideration to the 
case of region $S$ that is symmetric with respect to its vertical 
and horizontal axis. 
We assume that a pair of BPM electrodes is placed in the 
horizontal plane on the walls of such 2-axis symmetric chamber, 
and that the electrodes themselves are symmetric with respect to 
the horizontal plane. Then the signals induced on the right 
({\it R}) and left ({\it L}) electrodes by a pencil beam 
passing at location $\vec a=(x_0,y_0)$ do not change 
when $y_0 \leftrightarrow -y_0$ (i.e., they are even 
functions of $y_0$). Moreover, from the vertical symmetry, 
$L(x_0,y_0 ) = R(-x_0,y_0 )$. Using these properties, as well as 
the same trick $\partial _y^2 e_s  =  - \partial _x^2 e_s$ as 
above in the sum for derivatives of $e_0$, we obtain the 
difference-over-sum signal ratio of BPM signals in a rather 
general form:
\begin{eqnarray}
\lefteqn{ \frac{R - L}{R + L} = \frac{x_0 \partial_x e_0}
 {e_0} \times \left \{ \; 1 \; + \frac{1}{2}
 \frac{\partial_x^3 e_0}{\partial_x e_0}
 \left( \frac{x_0^2}{3} - y_0^2 + M_2 \right) \right. } \nonumber    
\\ && - \frac{1}{2}\frac{{\partial _x^2 e_0 }}{{e_0 }}
 \left( {x_0^2 - y_0^2 + M_2} \right) + \frac{1}{4}
 \left( \frac{{\partial _x^2 e_0 }}{{e_0 }} \right)^2
 \left( {x_0^2 - y_0^2 + M_2} \right)^2  \nonumber
\\ \lefteqn{ - \frac{1}{4}\frac{\partial_x^3 e_0}{\partial_x e_0}
     \frac{{\partial _x^2 e_0 }}{{e_0 }}
 \left( \frac{x_0^2}{3} - y_0^2 + M_2 \right)
  \left( {x_0^2 - y_0^2 + M_2} \right) }            \label{ratgen} 
\\ \lefteqn{ +  \frac{1}{24}\frac{{\partial_x^5 e_0 }}
 {\partial_x e_0}\left[ {\frac{x_0^4}{5} - 2x_0^2 y_0^2 + y_0^4 
 + 2M_2 \left( x_0^2 - 3y_0^2 \right) + M_4 } \right] } \nonumber 
\\ \lefteqn{ - \frac{1}{24}\frac{\partial_x^4 e_0}{e_0}
 \left[ x_0^4 - 6x_0^2 y_0^2 + y_0^4 + 6M_2 \left( x_0^2 - 
 y_0^2 \right) + M_4 \right] } \nonumber
\\ \lefteqn{ \left. + O\left( r_0^6/b^6,
  \sigma^6/b^6 \right) \right\} \ , } \nonumber
\end{eqnarray}
where the non-linearities are shown explicitly as powers of $x_0$ 
and $y_0$, and all beam-size corrections enter via the even moments 
$M_{2n}$ of the beam charge distribution, cf.\ Eq.\ (\ref{mexp2}).
One should note that in Eq.\ (\ref{ratgen}) we implicitly assume
values of the field $e_0$ and its derivatives averaged 
(or integrated, because we deal only with field ratios) over the
transverse extent of the right ({\it R}) electrode. It means that, 
for brevity, $\partial_x^n e_0, \ n=0,1,2,\ldots $ stands here for 
$\overline{\partial_x^n e_0} = 
1/\tau_R \int_R \partial_x^n e_0(\tau) d\tau$, 
where $\tau$ is a tangential length parameter of the electrode 
in its transverse cross section, and $\tau_R=\int_R d\tau$ is the 
electrode transverse width.

The general structure of non-linearities and beam-size corrections 
is rather clear in Eq.\ (\ref{ratgen}). It takes a relatively small 
effort to arrive to its particular case for the circular pipe, 
Eq.\ (\ref{ratcirc}). We just note that for the circular pipe
$\overline{e_0}=e_0=1/(2\pi b)$ and
$$
 \partial_x^n e_0=\frac{n!\cos{n\theta}}{\pi b^{n+1}} \ , 
 \ n=1,2,\ldots
$$
Averaging over the right electrode azimuthal extent, 
$-\phi/2 \le \theta \le \phi/2$, one gets
$$
 \overline{\partial_x^n e_0}=\frac{2(n-1)!}
 {\pi b^{n+1}} \frac{\sin{n\phi/2}}{\phi} \ , \ n=1,2,\ldots
$$
After that obtaining Eq.\ (\ref{ratcirc}) from (\ref{ratgen}) is
straightforward, taking into account the expressions for the
distribution moments $M_2,M_4$ of a double-Gaussian beam, cf.\ 
Eq.\ (\ref{mexpg}).

We conclude our study of the general case with a remark that the 
pencil beam field $e(\vec a,\vec b)$ and its derivatives are 
generally not easy to calculate, except for a few particular cases. 
Obviously, they include the case of a circular pipe where we know 
the explicit expression (\ref{fld0}) for $e(\vec a,\vec b)$. 
Another case where the eigenfunctions are simple and the sums 
in Eqs.\ (\ref{mexp2}) and (\ref{ratgen}) can be worked out 
relatively easy, is a rectangular chamber.

\section{Rectangular Chamber}

Let us consider a vacuum chamber with the cross section $S$ having 
a rectangular shape with width $w$ and height $h$. 
The orthonormalized eigenfunctions of the boundary problem 
(\ref{bound}) for region $S$ are
$$
e_{n,m} (x,y) = \frac{2}{{\sqrt {wh} }}
\sin \pi n\left( {\frac{1}{2} + \frac{x}{w}} \right)
\sin \pi m\left( {\frac{1}{2} + \frac{y}{h}} \right) \ ,
$$
where $ - w/2 \le x \le w/2$, $ - h/2 \le y \le h/2$, and 
$n,m=1,2,\ldots$. Summing up in Eq. (\ref{gexp}) for this case gives 
us the field produced by a pencil beam
\begin{eqnarray}
e(\vec r_0 ,\vec b) & = &
 \sum\limits_{m = 1}^\infty 
 \sin \pi m\left( \frac{h+y_0}{2h} \right)
 \sin \pi m\left( \frac{h+y_h}{2h} \right)  \nonumber
\\ & \times & \frac{ 2 \, \sinh \pi m\left[(w/2+ x_0)/h \right] } 
{h\, \sinh \left( \pi mw / h \right)}                   \label{fld0r}
\end{eqnarray}
at point $\vec b=(w/2,y_h)$ on the right side wall. Should we consider 
a left wall point instead, $\vec b=(-w/2,y_h)$, the only change in 
(\ref{fld0r}) would be the replacement $x \to - x$, see Sect.\ III 
for more general consideration of the symmetry. For points on top or 
bottom walls, one should exchange $w \leftrightarrow h,\;
x \leftrightarrow y$, and $y_h  \leftrightarrow x_w$ in Eq.\
(\ref{fld0r}). Unlike the circular pipe case, we are still left with 
a sum in Eq.\ (\ref{fld0r}), but the series is fast (exponentially) 
converging and convenient for calculations, e.g.\ see \cite{SK96,KGS}.
In particular, it is very easy here to calculate derivatives 
required in Eqs.\ (\ref{mexp2}) and 
(\ref{mexpg}-\ref{ratgen}): $\partial _x^2 e(\vec r,\vec b)$
is given by the same series (\ref{fld0r}), only with an extra factor
$(\pi m/h)^2$ in the sum. In fact, for the particular charge 
distributions (\ref{dblgs}) and (\ref{unifr}) considered above, it 
is simple enough to perform the integration (\ref{fldd}) directly 
using (\ref{fld0r}), which produces 
\begin{eqnarray}
\lefteqn{ E(\vec r_0 ,\vec b) = 
 \sum\limits_{m = 1}^\infty 
 \sin \pi m\left( \frac{h+y_0}{2h} \right)
 \sin \pi m\left( \frac{h+y_h}{2h} \right) }  \nonumber     
\\ && \times  \frac{ 2 \, \sinh \pi m\left( {\frac{w}{2h} 
 + \frac{x_0}{h}} \right)} {h\, \sinh \left( \pi mw / h \right)}
  f\left( {\frac{\pi m\sigma _y}{h}} \right)
  F\left( {\frac{\pi m\sigma _x}{w}} \right) .          \label{flddr}
\end{eqnarray}

The beam-size corrections in (\ref{flddr}) enter as the form-factors 
$f(z),F(z)$. For the double Gaussian charge distribution 
(\ref{dblgs}), the form-factors are $f(z) = \exp (-z^2/2)$, 
$F(z) = \exp (z^2/2)$, so that the correction factor in 
(\ref{flddr}) takes the form
$$
f\left( {\frac{{\pi m\sigma _y }}{h}} \right)
F\left( {\frac{{\pi m\sigma _x }}{w}} \right) = 
\exp \left[ {\left( {\frac{{\pi m}}{h}} \right)^2 
\frac{{\sigma _x^2  - \sigma _y^2 }}{2}} \right] \ .
$$
Obviously, for an axisymmetric beam with $\sigma_x = \sigma_y $ 
the argument of the exponent vanishes, and the exponent is equal to 
unity. As a result, the field (\ref{flddr}) of a finite-size 
axisymmetric beam will be exactly equal to that of a pencil beam,
Eq.\ (\ref{fld0r}). 

For the uniform rectangular distribution (\ref{unifr}),
the form-factors are $f(z) = \sin (z)/z$, $F(z) = \sinh (z)/z$, 
and the resulting correction factor is
$$
f\left( {\frac{{\pi m\sigma _y }}{h}} \right)
F\left( {\frac{{\pi m\sigma _x }}{w}} \right) = 
\frac{{\sin \left( {\pi m\sigma _y /h} \right)}}{{\pi m\sigma _y /h}}
 \frac{{\sinh \left( {\pi m\sigma _x /h} \right)}}
 {{\pi m\sigma _x /h}} \ .
$$
Expanding this expression in powers of $\sigma$ leads to the 
conclusion that the lowest beam-size corrections here have the 
order of $\sigma^4$, as we already know from Sect.\ III.

As for BPM signals, the simplest way is to use the general result 
(\ref{ratgen}). For two stripline BPM electrodes of width $h_1$ on 
side walls of a rectangular vacuum chamber $w \times h$, the 
difference over sum signal ratio, up to the 5th order, is
\begin{eqnarray}
\lefteqn{ \frac{R - L}{R + L}
 =  \pi \frac{x_0}{h} \frac{\Sigma_1}{\Sigma_0} 
  \times \left\{ \; 1 \; + \frac{\pi^2}{2h^2}
    \frac{\Sigma_3}{\Sigma_1} \! \left( \frac{x_0^2}{3} 
       - y_0^2  + M_2 \right)\right. }  \nonumber
\\ &&  -  \frac{\pi^2}{2h^2}\frac{\Sigma_2}{\Sigma_0}
  \left( {x_0^2  - y_0^2  + M_2 } \right)
 + \frac{\pi^4}{4h^4} \frac{\Sigma_2^2}{\Sigma_0^2}
   \left( {x_0^2 - y_0^2 + M_2 } \right)^2    \nonumber
\\ && - \frac{\pi^4}{4h^4}\frac{\Sigma_3}{\Sigma_1}
 \frac{\Sigma_2}{\Sigma_0} \! \left( \frac{x_0^2}{3} - y_0^2 
  + M_2 \right)\left( {x_0^2 - y_0^2 + M_2 } \right)    \label{ratr}   
\\ && + \frac{\pi^4}{24h^4}\frac{\Sigma_5}{\Sigma_1}
 \left[ \frac{x_0^4}{5} - 2x_0^2 y_0^2  + y_0^4  + 2M_2 
 \left(x_0^2 - 3y_0^2 \right) + M_4 \right]   \nonumber
\\ && - \frac{\pi^4}{24h^4}
 \frac{\Sigma_4}{\Sigma _0} \left[ x_0^4 - 6x_0^2 y_0^2 + y_0^4 
  + 6M_2 \left( x_0^2 - y_0^2 \right) + M_4 \right] \nonumber
\\ && \left. + O\left( r_0^6/b^6,
   \sigma^6/b^6 \right) \right\} \ , \nonumber
\end{eqnarray}
where $M_2,M_4$ are the moments of the beam charge distribution  
defined above, and
\begin{eqnarray*}
 \Sigma _{2n} & = & \sum\limits_{k = 0}^\infty 
 \left( 2k + 1 \right)^{2n} 
 \frac{\Phi \left( \pi \left( {k+1/2} \right) h_1 /h \right)}
{\cosh \left[ \pi \left( k+1/2 \right) w/h \right]} \ ;  
 \\  \Sigma _{2n + 1} & = & \sum\limits_{k = 0}^\infty  
 \left( {2k + 1} \right)^{2n + 1} 
 \frac{\Phi \left( \pi \left( k + 1/2 \right)h_1 /h \right)}
{\sinh \left[ \pi \left( k + 1/2 \right) w/h \right]} \ , 
\end{eqnarray*}
for $n=0,1,2,\ldots$. The sums above include one more form-factor,
$\Phi (z)= \sin{z}/z$, that accounts for the BPM electrode width. 
For narrow electrodes, when $h_1 \ll h$, it tends to 1.

Corrections (\ref{ratr}) are shown in Figs.\ 6-9 for a square 
chamber, $w=h$, and a BPM with very narrow electrodes, $h_1=h/100$
(in fact, results for $h_1=h/10$ are almost identical). Figure 6 
shows the BPM non-linearities for a pencil beam. Comparing it with 
the signal ratios for flat beams in Figs.\ 7-8, we notice a 
significant dependence of the signal ratio on the beam shape. 
Similar to Fig.\ 5, in Fig.\ 9 $S=(R - L)/(R + L)$ for a 
finite-size beam, while $S_0$ is the same ratio for a pencil beam, 
which is plotted in Fig.\ 6. Therefore, $S/S_0=1$ in Fig.\ 9 would 
mean that there was no correction due to a finite transverse beam 
size. As one can see, the beam-size corrections here are rather 
large, and they also depend noticeably on the beam vertical offset. 
The corrections range from about +3\% for $y=0$ (the chamber 
mid-plane) to less than 1\% for $y=h/8$ to about -(7-10)\% for 
$y=h/4$ (the beam is half-way to the top wall), in the case of 
$\sigma_x/w=0.1$, $\sigma_y/h=0.05$ shown in Fig.\ 9. 

These results are quite different from those for the 
wide-electrode BPM in a circular chamber (Sect.\ II), where the 
beam-size corrections were rather small. This is mainly because 
the considered square BPM has narrow electrodes, and not due to 
the different shape of its cross section.

\section{Conclusions}

Non-linearities and corrections due to a finite transverse beam size 
in beam fields and BPM signals are calculated for a homogeneous 
vacuum chamber in the case when either the wavelength of interest is 
longer than a typical transverse dimension of the chamber and/or the 
beam is ultra relativistic.
 
A general proof is presented that transverse beam-size corrections 
vanish in all orders for any azimuthally symmetric beam in an 
arbitrary chamber. One should emphasize that non-linearities are 
still present in this case; for a given chamber cross section, 
they depend only on the displacement of the beam center from the 
chamber axis. However, the non-linearities are the same for a 
finite-size axisymmetric beam and for a pencil beam (line source) 
with the same displacement. Having a non-symmetric transverse  
distribution of the beam charge results in additional (properly
beam-size) corrections. They tend to be minimal when the 
beam charge distribution is more symmetric. 

Explicit analytical expressions are derived for two particular cases 
--- circular and rectangular chamber cross section, as well as for 
the particular beam charge distributions --- double-Gaussian and 
uniform rectangular distribution.

While we have not discussed this subject in the present paper, the 
calculated corrections to beam fields can be directly applied in 
calculating beam coupling impedances produced by small discontinuities 
of the vacuum chamber using the methods of Refs.\ \cite{KGS,SK96}.

The author would like to acknowledge useful discussions with 
A.V.~Aleksandrov and M.M. Blaskiewicz.

\begin{figure}[htb]
\centerline{\psfig{figure=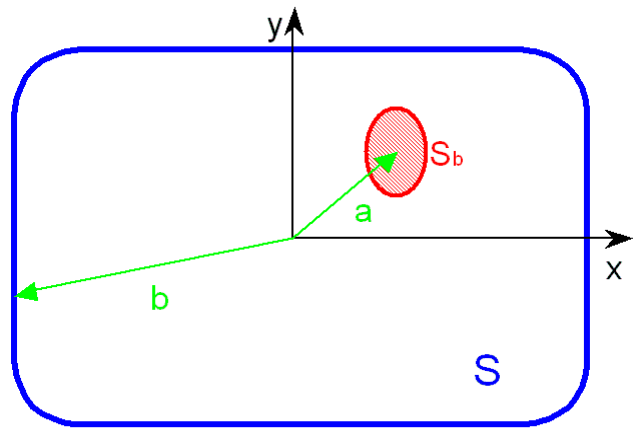,width=8.5cm}}
\caption{Transverse cross section of the vacuum chamber $S$ and 
of the beam $S_b$.} 
\end{figure}

\begin{figure}[htb]
\centerline{\psfig{figure=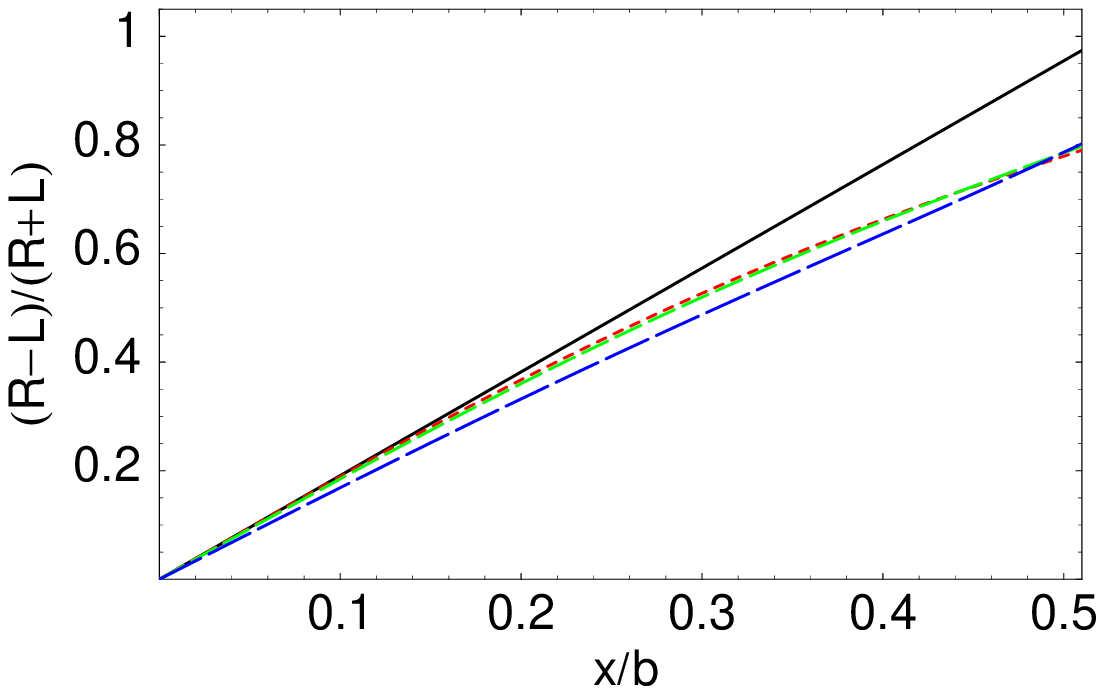,width=8.5cm}}
\caption{BPM signal ratio (\ref{ratcirc}) in a circular chamber  
versus beam center position $x/b$ for three vertical beam offsets 
$y/b=0,1/4,1/2$ (short-dashed, dashed, long-dashed) without 
beam-size corrections (pencil beam, $\sigma_x=\sigma_y=0$). 
Solid line shows the linear part of the BPM response.} 
\end{figure}

\begin{figure}[htb]
\centerline{\psfig{figure=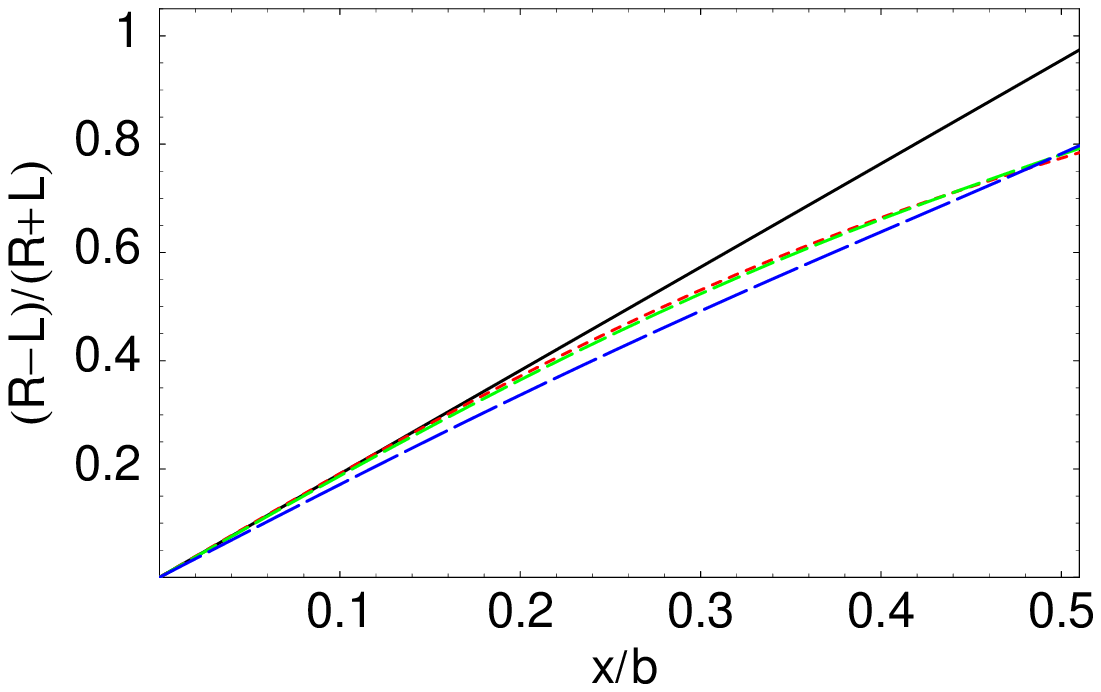,width=8.5cm}}
\caption{Same as Fig.\ 2, but with $\sigma_x/b=0.2$, $\sigma_y=0$.}
\end{figure}

\begin{figure}[htb]
\centerline{\psfig{figure=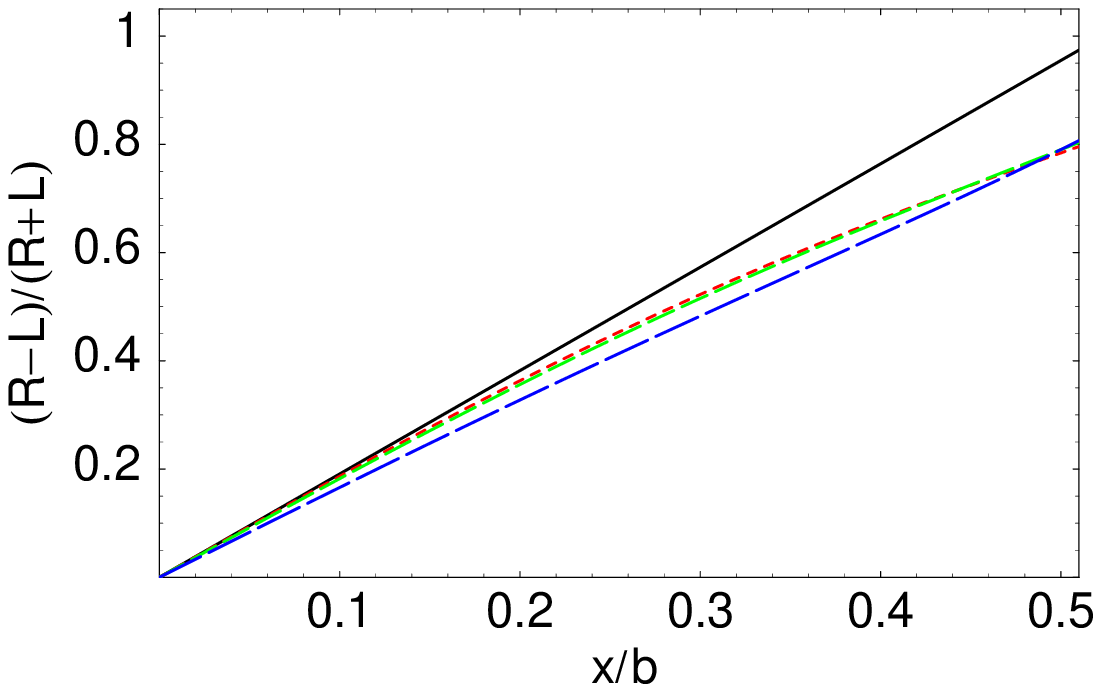,width=8.5cm}}
\caption{Same as Fig.\ 2, but with $\sigma_x=0$, $\sigma_y/b=0.2$.}
\end{figure}

\begin{figure}[htb]
\centerline{\psfig{figure=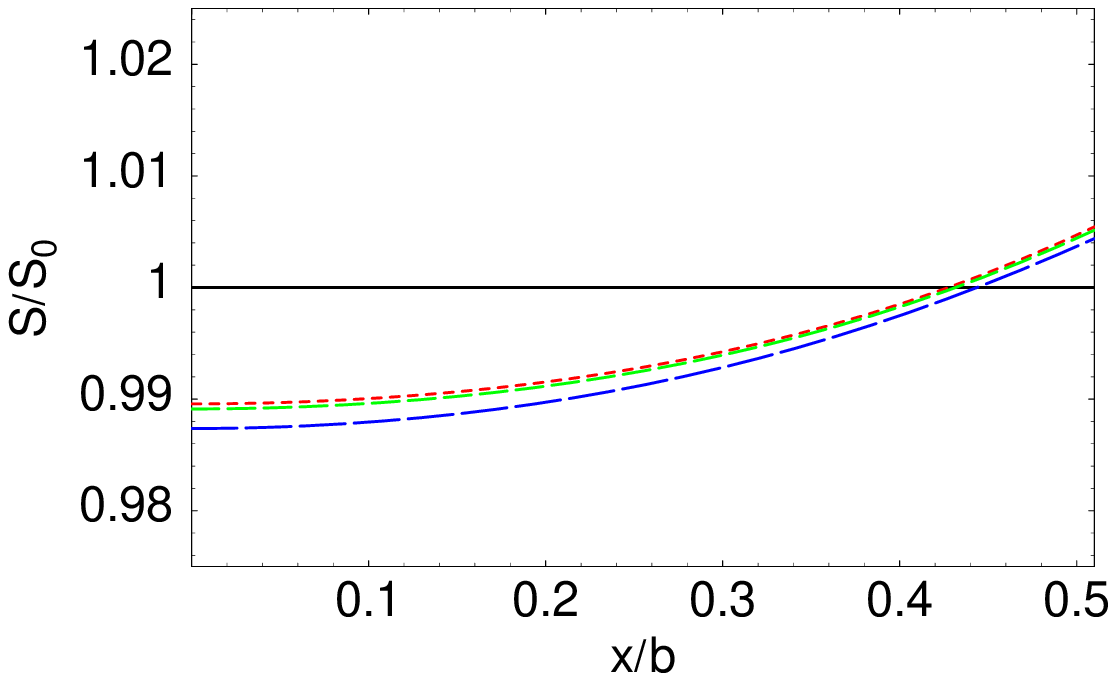,width=8.5cm}}
\caption{Relative magnitude of beam-size corrections in a circular 
chamber with $\sigma_x/b=0.1$, $\sigma_y/b=0.2$ for three vertical 
beam offsets $y/b=0,1/4,1/2$ (short-dashed, dashed, long-dashed). 
Here 1 corresponds to a pencil beam case, i.e.\ to one of the three 
curves in Fig.\ 2 for the corresponding beam vertical offset.}
\end{figure}

\begin{figure}[htb]
\centerline{\psfig{figure=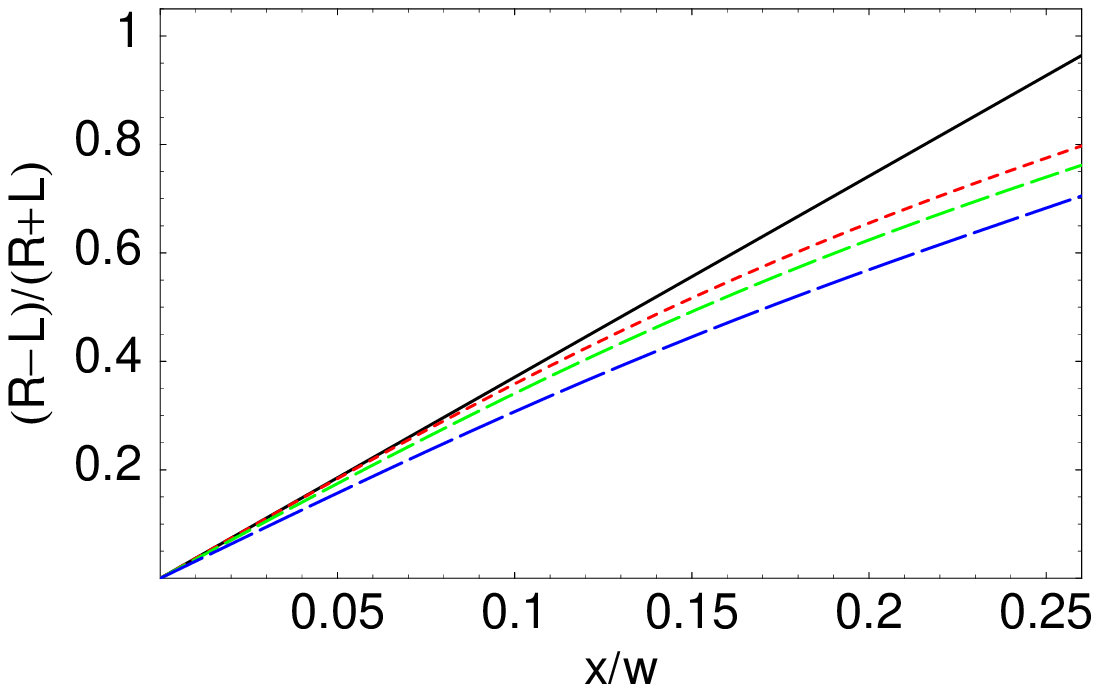,width=8.5cm}}
\caption{BPM signal ratio (\ref{ratr}) in a square chamber 
versus beam center position $x/w$ for three vertical beam offsets 
$y/h=0,1/8,1/4$ (short-dashed, dashed, long-dashed) without 
beam-size corrections (pencil beam, $\sigma_x=\sigma_y=0$). Solid 
line shows the linear part of the BPM response.} 
\end{figure}

\begin{figure}[htb]
\centerline{\psfig{figure=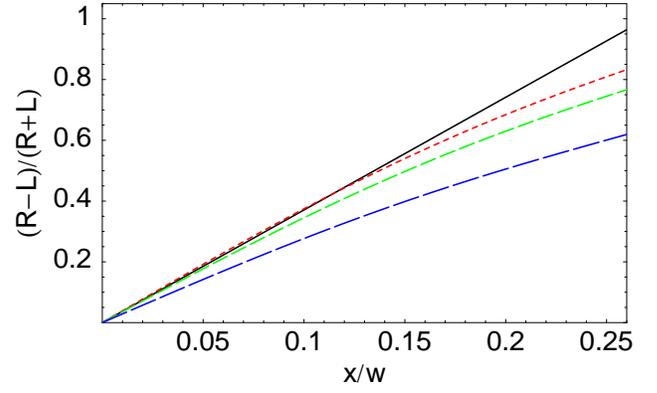,width=8.5cm}}
\caption{Same as Fig.\ 6, but with $\sigma_x/w=0.1$, $\sigma_y=0$.}
\end{figure}

\begin{figure}[htb]
\centerline{\psfig{figure=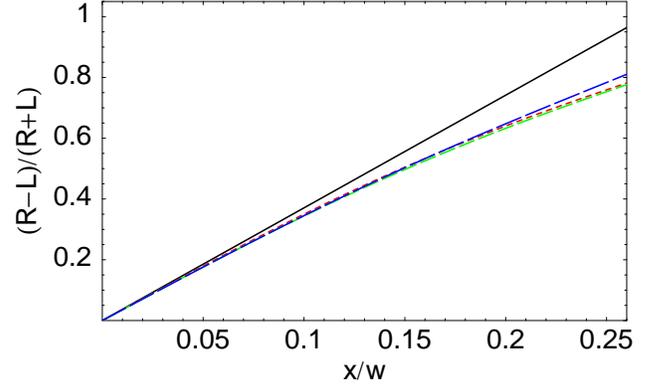,width=8.5cm}}
\caption{Same as Fig.\ 6, but with $\sigma_x=0$, $\sigma_y/h=0.1$.}
\end{figure}

\begin{figure}[htb]
\centerline{\psfig{figure=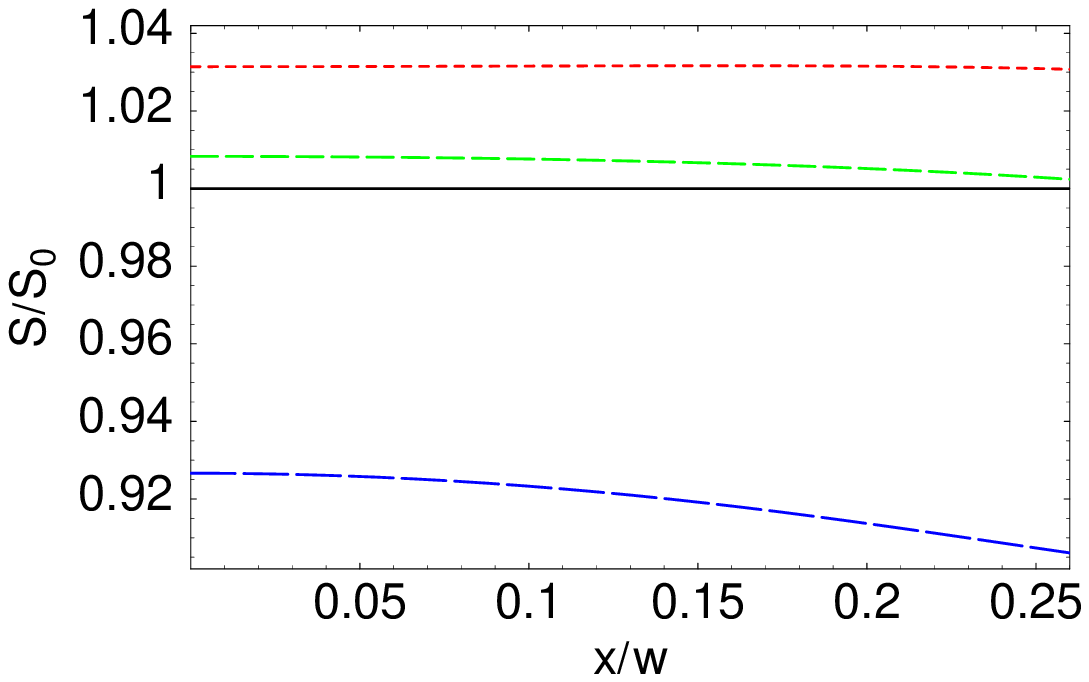,width=8.5cm}}
\caption{Relative magnitude of beam-size corrections in a square 
chamber with $\sigma_x/w=0.1$, $\sigma_y/h=0.05$ for three vertical 
beam offsets $y/h=0,1/8,1/4$ (short-dashed, dashed, long-dashed). 
Here 1 corresponds to a pencil beam case, i.e.\ to one of the three 
curves in Fig.\ 6 for the corresponding beam vertical offset.}
\end{figure}

\end{document}